\documentclass{ws-procs975x65}

\def \bx {{\bf x}}
\def \bp {{\bf p}}
\def \bn {{\bf n}}
\def \bm {{\overline{m}}}
\def \mycdot {{\!\,\cdot\!\,}}
\def \myacdot {{\hspace{-1.3pt}\cdot\!\,}}

\begin{document}

\title{POST-MINKOWSKIAN CLOSED-FORM HAMILTONIAN FOR GRAVITATING $N$-BODY SYSTEMS}

\author{TOM\'{A}\v{S} LEDVINKA$^*$, GERHARD SCH\"AFER$^{**}$ and JI\v{R}\'{I} BI\v C\'AK$^*$}

\address{$^*$Institute of Theoretical Physics\\
  Charles University, Prague\\
}

\address{$^{**}$Theoretisch-Physikalisches Institut\\
  Friedrich-Schiller-Universit\"at, Jena}

\begin{abstract}
The Hamiltonian for a system of relativistic bodies interacting by their
gravitational field is found in the post-Minkowskian approximation,
including all terms linear in the gravitational constant. It is given
in a surprisingly simple closed form as a function of canonical variables
describing the bodies only. The field is eliminated by solving inhomogeneous
wave equations, applying transverse-traceless projections, and using
the Routh functional. By including all special relativistic effects
our Hamiltonian extends the results described in classical textbooks
of theoretical physics. As an application, the scattering of relativistic
objects is considered.
\end{abstract}


\bodymatter\bigskip
Post-Newtonian (PN) approximation methods in general relativity
are based on the weak-field limit in which the metric is close to the
Minkowski metric and the assumption that the typical
velocity $v$ in a system divided by the speed of light is
very small. In post-Minkowskian (PM) approximation methods only the weakness
of the gravitational field is assumed but no assumption 
about slowness of motion is made. In the PM approximation we obtain\cite{LSB} 
the Hamiltonian for gravitationally interacting particles 
that includes all terms linear in gravitational constant $G$.
It thus yields PN approximations to {\it any} order in $1/c$ when terms linear
in $G$ are considered; and it can also describe
particles with ultrarelativistic velocities or with zero rest mass.

We use the canonical formalism of Arnowitt, Deser, and Misner
(ADM) \cite{ADM} where the independent degrees of freedom of the gravitational field are described
by $h_{ij}^{TT}$, the transverse-traceless part of $h_{ij}=g_{ij}-\delta_{ij}$
($h^{TT}_{ii}=0$, $h^{TT}_{ij,j}=0$, $i,j=1,2,3$), and by conjugate
momenta $c^3/(16\pi G) {\pi}^{ij\,TT}$. The field is generated by $N$ particles with rest
masses $m_a$ located at $\bx_a$, $a=1, ... N$, and with momenta $\bp_a$.
We start with the Hamiltonian\cite{S86} correct up to $G^2$ found by 
the expansion of the Einstein equations (the energy and momentum constraints) 
in powers of $G$ and by the use of suitable regularization procedures.
When we consider only terms linear in $G$ and put $c=1$ this Hamiltonian reads
\begin{align}
\label{HlinGS}
 H_{\rm lin}=&
\sum_a \bm_a - \frac{1}{2}G\sum_{a,b\ne a} \frac{\bm_a \bm_b }{ r_{ab} }
\left( 1+ \frac{p_a^2}{\bm_a^2}+\frac{p_b^2}{\bm_b^2}\right)
\\
\nonumber
&+ \frac{1}{4}G\sum_{a,b\ne a} \frac{1}{r_{ab}}\left( 7\, \bp_a\myacdot\bp_b + (\bp_a\myacdot\bn_{ab})(\bp_b\mycdot\bn_{ab})   \right)
-\frac{1}{2}\sum_a \frac{p_{ai}p_{aj}}{\bm_a}\,h_{ij}^{TT}(\bx=\bx_a)
\\\nonumber
\nonumber
&+\frac{1}{16\pi G}\int d^3x~ \left( \frac{1}{4} h_{ij,k}^{TT}\, h_{ij,k}^{TT} +\pi^{ij\,TT} \pi^{ij\,TT}\right)~,
\end{align}
where $\bm_a=\left( m_a^2+\bp^2_a \right)^\frac{1}{2}$,
$\bn_{ab} r_{ab} = \bx_a-\bx_b$, $|\bn_{ab}|=1$.
The equations of motion for particles are standard Hamilton equations, the Hamilton equations for the field read
\begin{equation}
 \dot {\pi}^{ij\,TT}~=~-16\pi G~\delta_{kl}^{TT\,ij} \frac{\delta H}{\delta h_{kl}^{TT}}
~,~~~
 \dot h_{ij}^{TT}~=~~16\pi G~\delta_{ij}^{TT\,kl} \frac{\delta H}{\delta {\pi}^{kl\,TT}}~;
\end{equation}
here the variational derivatives and the TT-projection operator
$\delta_{kl}^{TT\,ij} = \frac{1}{2}\left( \Delta_{ik}\Delta_{jl}+\Delta_{il}\Delta_{jk}-\Delta_{ij}\Delta_{kl}\right){\Delta^{-2}}$,
 $\Delta_{ij} = \delta_{ij}\Delta - \partial_i\,\partial_j$,  appear.
These equations imply the equations for the gravitational field
in the first PM approximation to be the wave equations with point-like sources $\sim\delta^{(3)}( \bx-\bx_a)$.
Since both the field and the accelerations $\dot \bp_a$
are proportional to $G$, the changes of the field due to the accelerations of particles are
of the order $O(G^2)$. Thus, in this approximation, 
wave equations can be solved assuming field to be generated by unaccelerated motion of particles,
i.e., it can be written as a sum of boosted static spherical fields:
\begin{equation}
\label{LiWi4h}
h_{ij}^{TT}(\bx) =
\delta_{ij}^{TT\,kl} \sum_b
\frac{4G}{\bm_b}
\frac{1}{|\bx-\bx_b|}
\frac{p_{bk}p_{bl}}{\sqrt {1-{\dot \bx_b}^2\sin^2 \theta_b}}~,
\end{equation}
where $\bx-\bx_a=\bn_a |\bx-\bx_a|$ and $\cos \theta_a={\bn_a\myacdot \dot \bx_a /|\dot \bx_a|}$.
Surprisingly, it is possible to convert the projection $\delta_{ij}^{TT\,kl}$ (which involves solving two Poisson 
equations) into an inhomogeneous linear second order ordinary differential equation and write 
\begin{align}
&h_{ij}^{TT}(\bx) ~=~ \sum_b
\frac{G}{|\bx-\bx_b|} \frac{1}{\bm_b}\frac{1}{y(1+y)^2}
\Big\{
\left[y\bp_b^2-(\bn_b\mycdot\bp_b)^2(3y+2)\right]\delta_{ij}
\\\nonumber&
+2\left[
1- \dot \bx_b^2(1 -2\cos^2 \theta_b)\right]{p_{bi}p_{bj}}
+\left[
\left( 2+y\right)(\bn_b\mycdot\bp_b)^2
\!-\!\left( 2+{3}y -2\dot \bx_b^2\right)\bp_b^2
\right]{n_{bi}n_{bj}}
\\\nonumber&
+2(\bn_b\mycdot\bp_b) \left(1-\dot \bx_b^2+2y\right) \left(n_{bi}p_{bj}+p_{bi}n_{bj}\right)
\Big\}
+O(\bm_b \dot \bx_b-\bp_b)G\!+\!O(G^2)~;
\label{unprojected_h}
\end{align}
here $y = y_b \equiv\sqrt {1-{\dot \bx_b}^2\sin^2 \theta_b}$ and we anticipate $O(\bm_b \dot \bx_b-\bp_b)\sim G$.

In the next step we use the Routh functional (see, e.g., Ref.\cite{DJS98})
\begin{equation}
R( \bx_a,\bp_a, h_{ij}^{TT}, \dot h_{ij}^{TT} ) =
H - \frac{1}{16\pi G}\int d^3x~  \pi^{TT\,ij}\, \dot h_{ij}^{TT}~,
\end{equation}
which is ``the Hamiltonian for the particles but the Lagrangian for the field.'' 
Since the functional derivatives of Routhian vanish if the field equations hold,
the (non-radiative) solution (\ref{LiWi4h}) can be substituted into the Routh functional 
without changing the Hamilton equations for the particles.
Using the Gauss's law, an integration by parts and similar standard steps 
(such as dropping out total time derivatives, i.e. a canonical transformation) 
and the explicit substitution for $h_{ij}^{TT}(\bx=\bx_a)$
we get the Hamiltonian for a $N$-particle gravitating system
in the PM approximation:
\begin{align}
\label{H1PM}
\nonumber
&H_{\rm lin}=
\sum_a \bm_a
+ \frac{1}{4}G\sum_{a,b\ne a} \frac{1}{r_{ab}}\left( 7\, \bp_a\myacdot\bp_b + (\bp_a\myacdot\bn_{ab})(\bp_b\mycdot\bn_{ab}) \right)
 - \frac{1}{2}G\sum_{a,b\ne a} \frac{\bm_a \bm_b }{ r_{ab}}
\\ &
\times\left( 1+ \frac{p_a^2}{ \bm_a^2}+\frac{p_b^2}{\bm_b^2}\right)
-\frac{1}{4}
G\sum_{a,b\ne a} \frac{1}{r_{ab}}
\frac{(\bm_a \bm_b)^{-1}}{ (y_{ba}+1)^2 y_{ba}}
\Bigg[
2\Big(2
(\bp_a\myacdot\bp_b)^2 (\bp_b\mycdot\bn_{ba})^2
\\\nonumber&
\!-\!2 (\bp_a\myacdot\bn_{ba}) (\bp_b\mycdot\bn_{ba}) (\bp_a\myacdot\bp_b) \bp_b^2
\!+\!(\bp_a\myacdot\bn_{ba})^2 \bp_b^4
\!-\!(\bp_a\myacdot\bp_b)^2 \bp_b^2
\Big ) \frac{1}{\bm_b^2}
+2 \Big[-\!\bp_a^2 (\bp_b\mycdot\bn_{ba})^2 
\\ \nonumber&
+ (\bp_a\myacdot\bn_{ba})^2 (\bp_b\mycdot\bn_{ba})^2 +
2 (\bp_a\myacdot\bn_{ba}) (\bp_b\mycdot\bn_{ba}) (\bp_a\myacdot\bp_b) +
(\bp_a\myacdot\bp_b)^2 - (\bp_a\myacdot\bn_{ba})^2 \bp_b^2\Big]
\\ \nonumber&
+
\Big[-3  \bp_a^2 (\bp_b\mycdot\bn_{ba})^2 +(\bp_a\myacdot\bn_{ba})^2 (\bp_b\mycdot\bn_{ba})^2
+8 (\bp_a\myacdot\bn_{ba}) (\bp_b\mycdot\bn_{ba}) (\bp_a\myacdot\bp_b) 
\\ \nonumber&
+ \bp_a^2 \bp_b^2 - 3 (\bp_a\myacdot\bn_{ba})^2 \bp_b^2 \Big]y_{ba}
\Bigg]~,~~~~~~~~~~~~~~~~~~~~~~~~~y_{ba} = \frac{1}{\bm_b} \sqrt{ m_b^2+ \left (\bn_{ba}\mycdot\bp_b\right)^2}~.
\end{align}

Since the PM approximation can describe ultrarelativistic or zero-rest-mass particles,
we calculated gravitational scattering of two such particles using the Hamiltonian (\ref{H1PM}).
If perpendicular separation $\bf b$ of trajectories ($|{\bf b}|$ is the impact parameter)
in the center-of-mass system ($\bp_1=-\bp_2\equiv\bp$) is used, $\bp{\mycdot}{\bf b}=0$,
we find, after evaluating a few simple integrals, that the exchanged momentum in the
system is given by
\begin{align}
\label{delta_p}
\Delta \bp &= -2\frac{{\bf b}}{{\bf b}^2} \frac{G}{|\bp|}
\frac{\bm_1^2 \bm_2^2}{\bm_1 +\bm_2 }
\left[
1+\left(\frac{1}{\bm_1^2}+\frac{1}{\bm_2^2}+\frac{4}{\bm_1\bm_2} \right)\bp^2
+\frac{\bp^4}{\bm_1^2 \bm_2^2 }
\right]~.
\end{align}
The quartic term is all that remains from the field part $h^{TT}_{ij}$ in agreement with Westpfahl\cite{W85}
who used a very different approach.

The Hamiltonian (\ref{H1PM}) can also describe a binary system with one massless and one massive particle orbiting around each other. This is not obvious: the second, fourth or even sixth-order PM approximation would not be able to describe massless test particles orbiting around a Schwarzschild black hole.

\bigskip
We acknowledge the support from SFB/TR7 in Jena,
from the Grant GA\v CR 202/09/0772 of the Czech Republic, and of Grants No LC 06014 and the MSM 0021620860
of Ministry of Education.

\end{document}